\documentclass{ws-ijmpa}
\usepackage[super,compress]{cite}
\usepackage{graphicx}
\usepackage{cancel}
\usepackage{cite}
\usepackage{slashed}
\usepackage{amsmath,bbm}
\usepackage{amssymb}
\usepackage{graphicx}
\usepackage[latin1]{inputenc}
\usepackage{mathrsfs}

\newcommand{\eps}{\varepsilon}
\newcommand{\bea}{\begin{eqnarray}}
\newcommand{\eea}{\end{eqnarray}}
\newcommand{\be}{\begin{equation}}
\newcommand{\ee}{\end{equation}}
\newcommand{\ba}{\begin{array}}
\newcommand{\ea}{\end{array}}

\def\gsim{\mathrel{\rlap{\lower4pt\hbox{\hskip1pt$\sim$}}
    \raise1pt\hbox{$>$}}}

\begin{document}

\markboth{Stefan Antusch and Oliver Fischer}{Probing the non-unitarity of the leptonic mixing matrix at the CEPC}

%
\catchline{}{}{}{}{}
%

\title{Probing the non-unitarity of the leptonic mixing matrix at the CEPC}

\author{Stefan Antusch\footnote{E-mail: stefan.antusch@unibas.ch} $^{1\,2}$ }

\author{Oliver Fischer\footnote{E-mail: oliver.fischer@unibas.ch} $^1$ }

\address{$^1$ Department of Physics, University of Basel,\\
Klingelbergstr. 82, CH-4056 Basel, Switzerland}

\address{$^2$ Max-Planck-Institut f\"ur Physik (Werner-Heisenberg-Institut),\\
F\"ohringer Ring 6, D-80805 M\"unchen, Germany}
\maketitle


\begin{abstract}
The non-unitarity of the leptonic mixing matrix is a generic signal of new physics aiming at the generation of the observed neutrino masses. We discuss the Minimal Unitarity Violation (MUV) scheme, an effective field theory framework which represents the class of extensions of the Standard Model (SM) by heavy neutral leptons, and discuss the present bounds on the non-unitarity parameters as well as estimates for the sensitivity of the CEPC, based on the performance parameters from the preCDR. 

\end{abstract}


\section{Introduction}	
The experimental results on neutrino oscillations have unambiguously shown that {\it at least} two of the three known SM neutrinos are massive. However, the mass generating mechanism is still unknown and constitutes one of the great open questions in particle physics. At present, no experimental evidence exists to tell us which among the various possible extensions of the SM to realise the observed neutrino masses is realised in nature.  

One of the best-motivated extensions of the SM to generate the masses for the SM neutrinos consists in adding sterile neutrinos  to the SM particle content, which are often also referred to as ``gauge singlet fermions'', ``heavy neutral leptons'' or simply ``heavy'' or ``right-handed'' neutrinos. 

When the sterile neutrinos are much heavier than the energy scale of a given experiment, only the light neutrinos propagate, which implies that the experiments are sensitive to an ``effective leptonic mixing matrix''. This effective leptonic mixing matrix is given by a submatrix of the full unitary leptonic mixing matrix, and thus is in general not unitary. The new physics effects within this class of SM extensions can be described in a model independent way by the effective theory framework of the Minimal Unitarity Violation (MUV) scheme\cite{Antusch:2006vwa}. 

In this article, we discuss the expected sensitivity of the Circular Electron Positron Collider (CEPC) and other present and future experiments for probing the non-unitarity of the leptonic mixing matrix in the MUV scheme, updating some of our results\cite{Antusch:2014woa} to account for the performance parameters from the CEPC preCDR.

\section{Origin of non-unitarity of the leptonic mixing matrix}
\label{sec:MUV}
A very general effective description of leptonic non-unitarity in scenarios with an arbitrary number of sterile neutrinos is given by the Minimal Unitarity Violation (MUV) scheme\cite{Antusch:2006vwa}. It describes the generic situation that the left-handed neutrinos mix with other neutral fermionic fields that are much heavier than the energy scale where the considered experiments are performed. 

Adding $n$ sterile neutrinos to the SM, the resulting generalized ``seesaw extension of the SM'' is given by 
\be\label{eq:The3FormsOfNuMassOp}
\mathscr{L} = \mathscr{L}_\mathrm{SM} -\frac{1}{2} \overline{N_\mathrm{R}^I} M^N_{IJ} N^{c\, J}_\mathrm{R} -(y_{\nu_\alpha})_{I\alpha}\overline{N_\mathrm{R}^I} \widetilde \phi^\dagger
L^\alpha+\mathrm{H.c.}\; ,
\ee
where the  $N_\mathrm{R}^I$ ($I=1,...,n$) are sterile neutrinos, $\tilde \phi := i \tau_2 \phi^*$ and $L^\alpha$ is the lepton doublet.  $\phi$ denotes the SM Higgs doublet field, which breaks the electroweak (EW) symmetry spontaneously by acquiring a vacuum expectation value $v_\mathrm{EW}=246.44$ GeV in its neutral component. 

When the mass scale of the sterile neutrinos, denoted by $M$ in the following, is much larger than $v_\mathrm{EW}$, the $N_\mathrm{R}^I$ can be integrated out of the theory below $M$ and an effective theory emerges which contains the effective dimension 5 and dimension 6 operators $\delta{\cal L}^{d=5} $ and $\delta{\cal L}^{d=6} $. The SM extended by these two effective operators defines the MUV scheme.
 
The operator $\delta{\cal L}^{d=5}$ generates the masses of the light neutrinos after electroweak symmetry breaking, and is given by
\bea
\delta{\cal L}^{d=5} = \frac{1}{2}\, c_{\alpha \beta}^{d=5} \,\left( \overline{L^c}_{\alpha} \tilde \phi^* \right) \left( \tilde \phi^\dagger \, L_{ \beta} \right) + \mathrm{H.c.} \;,
\label{eq:dim5op}
\eea
The second effective operator
\bea
\delta{\cal L}^{d=6} = c^{d=6}_{\alpha \beta} \, \left( \overline{L}_{\alpha} \tilde \phi \right) i \cancel{\partial} \left(\tilde \phi^\dagger L_{ \beta} \right)
\label{eq:dim6op}
\eea
generates a contribution to the kinetic terms of the neutrinos (and only the neutrinos) after EW symmetry breaking. Canonically normalising the kinetic terms in general involves a transformation of the neutrino fields by a non-unitary matrix and leads to a non-unitary (effective) leptonic mixing matrix (see, e.g., references\cite{DeGouvea:2001mz,Broncano:2002rw,Antusch:2006vwa}).

The coefficient matrix in the definition of $\delta{\cal L}^{d=5}$, (cf.\ eq.~(\ref{eq:dim5op}),) can be connected to the parameters of the seesaw extension of the SM by:
\bea
c_{\alpha \beta}^{d=5} = (y_\nu^T)_{\alpha I} (M_N)_{IJ}^{-1} (y_\nu)_{J\beta}\:.
\eea
After EW symmetry breaking, $\delta{\cal L}^{d=5}$ generates the mass matrix of the light neutrinos: 
\bea\label{Eq:d5AndMnu}
(m_\nu )_{\alpha\beta}= - \frac{v_\mathrm{EW}^2}{2} c_{\alpha\beta}^{d=5}\;,
\eea 
which corresponds to the usual (generalised) seesaw formula. The coefficient matrix for the dimension 6 operator, i.e.\ $c_{\alpha \beta}^{d=6}$ in eq.~(\ref{eq:dim6op}), is related to the sterile neutrino parameters by (cf.\ reference [\refcite{Broncano:2003fq}])
\bea
c_{\alpha \beta}^{d=6} =  \frac{y_{\nu_\alpha}^* y_{\nu_\beta}}{M^2}\;.
\label{eq:d=6coeffs}
\eea
After EW symmetry breaking and the canonical normalisation of the neutrino kinetic terms, the unitary mixing matrix in the lepton sector is modified to the (effective) non-unitary leptonic mixing matrix ${\cal N}$. ${\cal N}$ is related to $c_{\alpha \beta}^{d=6}$ via: 
\be
({\cal N N}^\dagger)^{-1}_{\alpha\beta}  - {1}_{\alpha\beta} = \frac{v^2_\mathrm{EW}}{2} c_{\alpha \beta}^{d=6} \:.
\label{eq:PMNS}
\ee
One can use the Hermitean matrix $\eps$ (with small entries) to parametrise the deviation of the leptonic mixing matrix $\cal  N$ from unitarity by defining 
\be
({\cal N N}^\dagger)_{\alpha\beta}  = {1}_{\alpha\beta} + \eps_{\alpha \beta} \:.
\ee
The matrix elements $\eps_{\alpha \beta}$ are related to $c_{\alpha \beta}^{d=6}$ by
\be\label{Eq:EpsAndNonU}
\eps_{\alpha \beta} = - \frac{v^2_\mathrm{EW}}{2} c_{\alpha \beta}^{d=6} \;,
\ee
up to higher order terms in the small elements $\eps_{\alpha \beta}$.

In the MUV scheme, the charged and neutral EW currents are modified as:
\be
j_\mu^\pm =  \bar \ell_\alpha \gamma_\mu {\cal N}_{\alpha j} \nu_j + \mbox{H.c.}\,, \quad j_\mu^0 = \bar \nu_i \,({\cal N}^\dagger {\cal N})_{ij} \,\gamma_\mu \nu_j\:,
\label{eq:weakcurrents}
\ee
where $\ell_\alpha$ denote the charged leptons and the $\nu_i,\,\nu_j$ denote the light neutrino mass eigenstates ($i,j \in\{1,2,3\}$). These modifications can lead to various observable effects, which allow to test the non-unitarity of the leptonic mixing matrix experimentally, as we will now discuss.

\section{Electroweak Precision Observables}
Due to the modifications of the EW interactions, the presence of sterile neutrinos changes the theory predictions for the Electroweak Precision Observables (EWPOs). For calculating the prediction in the MUV scheme we make use of the high precision of the most accurately measured\cite{Agashe:2014kda} parameters:
\be
\begin{split}
\alpha(m_z)^{-1} & = 127.944(14) \:, \\
G_F & = 1.1663787(6)\, \times 10^{-5} \text{GeV}^{-2} \:, \\ 
m_Z & = 91.1875(21)\: , 
\end{split}
\ee
where $m_Z$ denotes the $Z$ pole mass, $\alpha$ the fine structure constant and $G_F$ the Fermi constant. 

The parameters $\alpha$ and $m_Z$ are not modified in the presence of sterile neutrinos (i.e.\ within MUV). The Fermi constant $G_F$, however, is measured from muon decay, which is sensitive to the charged current interactions (cf.\ eq.\ (\ref{eq:weakcurrents})). The experimentally measured quantity is the muon decay constant $G_\mu$, which, in the MUV scheme, has the following tree level relation to $G_F$:
\be
 G_\mu^2 = G_F^2({\cal NN}^\dagger)_{\mu \mu}({\cal NN}^\dagger)_{ee}\;,
\label{eq:GF}
\ee
where we have assumed that $M > m_\mu$ and ${\cal N}$ is the non-unitary leptonic (PMNS) mixing matrix. This is very important since the Fermi constant enters the theory predictions for most EWPOs. For a list of modifications of the EWPOs, see reference [\refcite{Antusch:2014woa}] and the references therein.

\section{Universality tests and low energy precision observables}
In addition to the EWPOs, also the lepton universality tests from $W$ decays at the CEPC will be very powerful probes of leptonic non-unitarity. Furthermore, various low energy precision observables are very relevant:\cite{Antusch:2014woa}

\begin{itemize}
\item The lepton universality observables are given by the ratios of decay rates: $R^X_{\alpha\beta}=\Gamma^X_\alpha/\Gamma^X_\beta$, where $\Gamma^X_\alpha$ denotes a  decay width including a charged lepton $\ell_\alpha$ and a neutrino. In the SM, $R^X_{\alpha\beta}=1$ holds for all $\alpha,\beta,X$, which is generally called lepton universality. 
In our analysis, we consider the low energy measurements of $\pi$, $\mu$, $\tau,$ $K$ and $W$ decays from experiments, which are very precise probes of the charged current as defined in eq.~\eqref{eq:weakcurrents}. 

\item Strong constraints on the non-unitarity of leptonic mixing can also be obtained from measurements of rare lepton flavour violating processes, in particular from the searches for charged lepton decays $\ell_\rho\to \ell_\sigma \gamma$. The presently strongest constraint comes from the MEG collaboration\cite{Adam:2013mnn} with the upper bound on the process $\mu \to e \gamma \leq 5.7 \times 10^{-13}$. Such experiments however probe only the off-diagonal non-unitarity parameters 

\item The very precise measurements of meson decays are used to determine the unitarity of the first row of the CKM matrix. The unitarity condition on the CKM matrix implies strong bounds on the modification of the weak currents in eq.~\eqref{eq:weakcurrents} and thus on the non-unitarity parameters.

\item The modified theory prediction of the weak mixing angle can be constrained by the results from the NuTeV experiment, which measured deep inelastic scattering of a neutrino beam off a nucleon. The initially significant deviation\cite{Zeller:2001hh} from the LEP measurement has been reinvestigated (cf.\ Ref.\ [\refcite{Bentz:2009yy}]) with the effect that the tension with other precision data was removed. 

\item In our analysis, we furthermore include low energy measurements of parity violation from the weak currents at energies far below the $Z$ boson mass. This approach can allow to measure the weak mixing angle with a precision below the percent level, see e.g.\ the reviews\cite{Erler:2004cx,Kumar:2013yoa} for an overview. 
Currently, the best measured values for the weak mixing angle at low energies come from the determination of the weak charge of Caesium\cite{Dzuba:2012kx} and of the proton\cite{Nuruzzaman:2013bwa}, and the parity violating asymmetry in M$\o$ller scattering\cite{Kumar:2013yoa}.

\end{itemize}

\subsection{Present constraints from a global fit}
We performed a Markov Chain Monte Carlo fit of the relevant six non-unitarity parameters in the Minimal Unitarity Violation (MUV), including the correlations between the observables.\footnote{We note that the phases of the complex off-diagonal non-unitarity parameters can be probed by future neutrino oscillation experiments \cite{FernandezMartinez:2007ms,Antusch:2009pm}. Also note that within the MUV scheme, the diagonal non-unitarity parameters are real and $\le 0$.}
Our global analysis yields the following highest posterior density (HPD) intervals at 68\% ($1\sigma$) Bayesian confidence level (CL):\cite{Antusch:2014woa}
\be
\begin{array}{ccl} \epsilon_{ee}  & = &-0.0012 \pm 0.0006 \\ | \epsilon_{\mu\mu} | & < & 0.00023 \\ \epsilon_{\tau\tau} & = & -0.0025 \pm 0.0017 \end{array} \qquad  \begin{array}{ccl} | \epsilon_{e\mu} | & < & 0.7 \times 10^{-5} \\ | \epsilon_{e\tau} | & < & 0.00135 \\ | \epsilon_{\mu\tau} | & < & 0.00048 \,,\end{array}
\label{MUVresult}
\ee
We remark that the moduli of the off-diagonal elements are generally restricted by the triangle inequality which implies\cite{Antusch:2008tz}:
\be
|\eps_{\alpha \beta}| \le \sqrt{|\eps_{\alpha \alpha}| |\eps_{\beta \beta}|}\:.
\label{eq:triangleineq}
\ee
We found that the best fit points for the off-diagonal $\eps_{\alpha\beta}$ and for $\eps_{\mu\mu}$ are at zero.
At 90\% CL, the constraints on the non-unitary PMNS matrix are:
\be
\left| {\cal NN}^\dagger \right| = \left( \begin{array}{ccc} 0.9979 - 0.9998 & < 10^{-5} & < 0.0021 \\
< 10^{-5} & 0.9996 - 1.0 & < 0.0008 \\
< 0.0021 & < 0.0008 & 0.9947 - 1.0 \end{array}\right)\,.
\ee
Our result show that non-zero $\eps_{ee}$ and $\eps_{\tau\tau}$ improve the fit to the data, while the best-fit value for $\eps_{\mu\mu}$ is zero. We note, that negative $\eps_{\alpha\alpha}$ were imposed as prior in the analysis (as required in the MUV scheme). 

In our fit, it turned out that the experimental bounds on $\eps_{e \tau}$ and $\eps_{\mu \tau}$ are comparable to those from eq.~\eqref{eq:triangleineq}. It is interesting to remark that a future observation of rare tau decays beyond the level allowed by the triangle inequality, or a strong experimental indication of a positive $\eps_{\alpha\alpha}$, has the potential to rule out the MUV scheme.

\section{Improved sensitivity to leptonic non-unitarity at the CEPC}
\label{sec:improvementsMUV}
As we will now discuss, the CEPC would provide excellent sensitivity to the non-unitarity of the leptonic mixing matrix, in particular via improved measurements of the Electroweak Precision Observables (EWPOs) and via universality tests using the leptonic decays of the large W boson sample. 

\subsection{Sensitivity via EWPOs}
The modifications of the theory predictions for the EWPOs depend on the sum $|\eps_{ee}|+|\eps_{\mu \mu}|$ and on $|\eps_{\tau\tau}|$. 
In order to estimate the possible sensitivity of the CEPC, we show the estimated precision for the EWPOs in the table below, based on the preCDR. Note that we consider the projected systematic uncertainties rather than the statistical ones, which could be much smaller. The theory uncertainties are set to zero unless stated otherwise. 

\begin{center}
\begin{tabular}{|l|c|c|c|}
\hline
Observable & LEP precision & from CEPC preCDR 
\\
\hline\hline
$M_W$ [MeV] & 33 & 3 
\\
$\sin^2 \theta_W^{\rm eff}$  & 0.07\% & 0.01\% 
 \\
$R_b$ & 0.3\% & 0.08\% 
 \\
$R_c$ & 0.3\% & 0.07\% \\
$R_{inv}$ & 0.27\% & 8.9$\times 10^{-4}$ 
 \\
$R_\ell$ & 0.1\% & 0.1\% 
\\
$\Gamma_\ell$ & 0.1\% & 0.1\% 
\\
$\sigma_h^0$ [nb]  & 8.9 $\times 10^{-4}$ & 1 $\times 10^{-4}$ \\
\hline
\end{tabular}
\end{center}

\begin{figure}
\begin{center}
\includegraphics[scale=0.55]{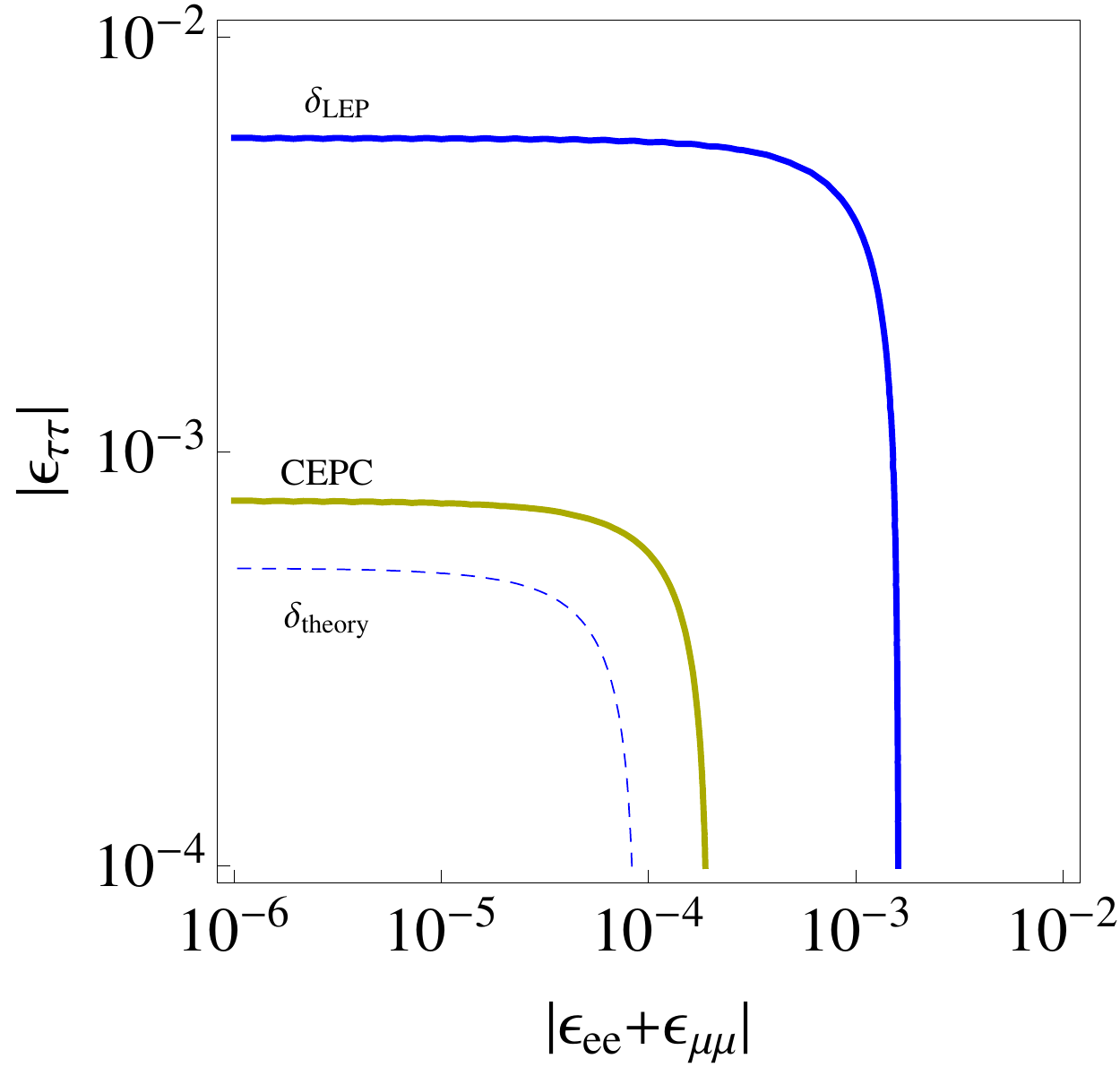}  %
\end{center}
\caption{Exclusion limits on the $|\eps_{\tau\tau}|$ and  $|\eps_{ee}|+|\eps_{\mu \mu}|$ at 90\% confidence level (updated version of the figure from reference\cite{Antusch:2014woa}). The solid blue line denotes the present experimental precision of the EWPOs from LEP, while the dashed blue line denotes the present theory uncertainty. The solid yellow line corresponds to the estimated precision of the CEPC.}
\label{fig:CEPC}
\end{figure}

The estimated sensitivity for the (combination of) non-unitarity parameters $|\eps_{ee}|+|\eps_{\mu \mu}|$ and $|\eps_{\tau\tau}|$ is shown in Fig.~\ref{fig:CEPC}, at the 90\% confidence level, by the solid yellow line. 
The solid blue line denotes the present constraints from LEP and the dashed blue line denotes the limit from the present theory uncertainty. The sensitivity of the CEPC to $|\eps_{ee}|+|\eps_{\mu \mu}|$ of $\sim 2\times 10^{-4}$ can be translated via eq.~(\ref{eq:d=6coeffs}) into a mass for the ``sterile'' or ``right-handed'' neutrino of $\sim 12$ TeV, for Yukawa couplings of order one.

\subsection{Sensitivities via lepton universality tests}

In order to estimate the sensitivities of the CEPC and other planned future experiments to leptonic non-unitarity via universality tests, we use the following performance parameters:

\begin{center}
\begin{tabular}{|l|cccc|}
\hline
Observable & $R^\ell$ & $R^\pi$ & $R^K$ & $Br(W\to\ell \nu)$  \\
\hline
Precision & 0.001 & 0.001 & 0.004 & 0.0003  \\
Experiment & Tau factories\cite{Biagini:2013fca} & TRIUMF\cite{AguilarArevalo:2010fv}, PSI\cite{Pocanic:2003pf} & NA62\cite{Goudzovski:2010uk} & CEPC\cite{CDR}  \\
\hline
\end{tabular}
\end{center}

\begin{figure}
\begin{center}
\includegraphics[scale=0.5]{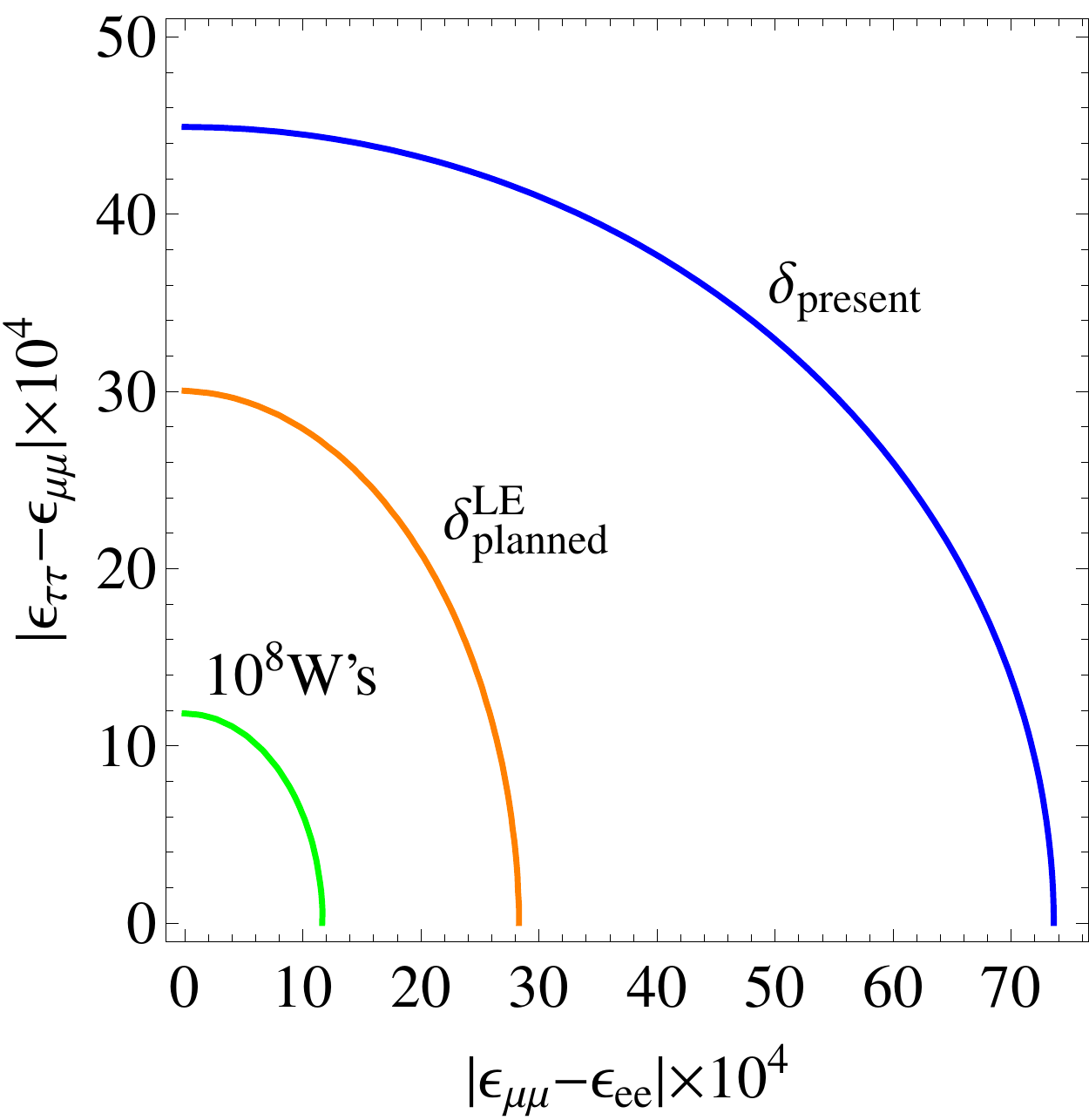}
\end{center}
\caption{Exclusion limits at 90\% confidence level\cite{Antusch:2014woa}. The blue line denotes the present experimental exclusion limits, the orange line represents the estimated improvements\cite{Mammei:2012ph,AguilarArevalo:2010fv,Pocanic:2003pf,Goudzovski:2010uk,Biagini:2013fca} in the low energy sector. The green line includes the estimated improvement from $W\to \ell_\alpha$ decays at the CEPC.}
\label{fig:LEprecision}
\end{figure}

The prediction for the universality observables in the MUV scheme depends on the differences between the diagonal non-unitarity parameters. We choose the following differences as parameters:
\be
\Delta_{\tau\mu} := \eps_{\tau\tau} - \eps_{\mu\mu}\, \qquad \text{and}\qquad \Delta_{\mu e} := \eps_{\mu\mu} - \eps_{ee}\,.
\ee
The exclusion sensitivity contours for the two combinations of non-unitarity parameters at 90 \% CL is shown in fig.~\ref{fig:LEprecision}. 
The blue line represents the current constraints, while the improvements by the planned low energy experiments are shown by the orange line. The estimated sensitivity of the CEPC (with $10^8$ W decays), combined with the future low energy experiments, is represented by the green line.

\subsection{Expected improvements from charged lepton flavour violation and neutrino oscillation experiments}
In the fit to the present data, the experimental constraint on the charged lepton flavour violating  (cLFV) decay $\mu \to e \gamma$ from MEG\cite{Adam:2013mnn} drives the strong bound on the non-unitarity parameter $|\eps_{e\mu}|$. 
Future tests of cLFV will provide a substantial improvement in precision. For example, the sensitivity of the PRISM/PRIME project\cite{Barlow:2011zza} and a Mu2e upgrade\cite{Knoepfel:2013ouy} could achieve a sensitivity of $|\eps_{e\mu}| \sim 3.6 \times10^{-7}$. 

Furthermore, the precision to the branching ratio for the cLFV tau decay $\tau \to e \gamma$ is expected to improve to $10^{-9}$ at SuperKEKB\cite{Akeroyd:2004mj}, which leads to an improved sensitivity of $|\eps_{e\tau}| \sim 1.5 \times10^{-3}$. 
Finally, we remark that a sensitivity to the phases of the parameters could be achieved in neutrino oscillation experiments, as discussed in Refs.\ [\refcite{FernandezMartinez:2007ms,Antusch:2009pm}] within the MUV scheme.

\section{Discussion and Conclusions}
\label{sec:conclusions}
In this article, we have reviewed and partly updated the results from our previous work \cite{Antusch:2014woa} where we have investigated the potential of future lepton colliders, such as the CEPC, for testing the non-unitarity of the leptonic mixing matrix. As framework we considered the Minimal Unitarity Violation (MUV) scheme, which is an effective field theory description of extensions of the SM by ``heavy'' sterile neutrinos. 

For ``heavy'' sterile neutrinos, i.e.\ with masses much larger than the EW scale, we presented the results for the present constraints on the non-unitarity parameters from a global fit in the MUV scheme. Although the present electroweak precision data favours sterile neutrinos with active-sterile mixing $|\theta_e|^2 = {\cal O}(10^{-3})$ at the $2\sigma$
 level, which might be a first hint at the existence of sterile neutrinos, we rather view the constraints on the non-unitarity parameters at the 90\% Bayesian confidence level as the main conclusion from the fit to the present data.
 
 Future experiments have the potential to push the search for new physics (such as sterile neutrinos) via the non-unitarity of the leptonic mixing matrix to a new level. A very high sensitivity to $|\eps_{ee}|+|\eps_{\mu \mu}|$ and $|\eps_{\tau \tau}|$ down to $ \sim 2 \times 10^{-4}$ and $ \sim 10^{-3}$, respectively, can be achieved via the precision measurements of the EWPOs at the CEPC. Also the universality tests in $W$ decays at the CEPC would be very powerful, and can yield a sensitivity to the differences of the diagonal non-unitarity parameters down to $ \sim  10^{-3}$. Furthermore, complementary experiments, for instance on cLFV and on neutrino oscillations, could probe very sensitively the off-diagonal flavour-violating non-unitarity parameters.

In summary, the CEPC would be a powerful experiment to probe the non-unitarity of the leptonic mixing matrix and thereby search for new physics beyond the SM (such as sterile neutrinos) needed to explain the observed masses of the light neutrinos.

\subsection*{Acknowledgements}
This work was supported by the Swiss National Science Foundation. We thank the organisers of the IAS Program on the Future of High Energy Physics in Hong Kong for their hospitality.

\bibliographystyle{unsrt}

\end{document}